# MATI: A GPU-Accelerated Toolbox for Microstructural Diffusion MRI Simulation and Data Fitting with a User-Friendly GUI

Junzhong Xu[1,2,3,4*], Sean P. Devan[1], Diwei Shi[5], Adithya Pamulaparthi[6], Nicholas Yan[7], Zhongliang Zu[1,2], David S. Smith[1,2,3,4], Kevin D. Harkins[1,2], John C. Gore[1,2,3,4], Xiaoyu Jiang[1,2]

[1] Institute of Imaging Science, Vanderbilt University Medical Center, Nashville, Tennessee.
[2] Department of Radiology and Radiological Sciences, Vanderbilt University Medical Center, Nashville, Tennessee.
[3] Department of Biomedical Engineering, Vanderbilt University, Nashville, Tennessee.
[4] Department of Physics and Astronomy, Vanderbilt University, Nashville, Tennessee.
[5] Tsinghua University, Beijing, China.
[6] Department of Electrical and Computer Engineering, Vanderbilt University, Nashville, Tennessee.
[7] High School, Knoxville, Tennessee.

[*] **Corresponding author**: Address: Institute of Imaging Science, Department of Radiology and Radiological Sciences, Vanderbilt University Medical Center, 1161 21st Avenue South, AA 1105 MCN, Nashville, TN 37232-2310, United States. Fax: +1 615 322 0734.
E-mail address: junzhong.xu@vanderbilt.edu (Junzhong Xu)

**Abstract**

MATI (Microstructural Analysis Toolbox for Imaging) is a versatile MATLAB-based toolbox that combines both simulation and data fitting capabilities for microstructural dMRI research. It provides a user-friendly, GUI-driven interface that enables researchers, including those without programming experience, to perform advanced MRI simulations and data analyses. For simulation, MATI supports arbitrary microstructural modeled tissues and pulse sequences. For data fitting, MATI supports a range of fitting methods—including traditional non-linear least squares, Bayesian approaches, machine learning, and dictionary matching methods—allowing users to tailor analyses based on specific research needs. Optimized with vectorized matrix operations and high-performance numerical libraries, MATI achieves high computational efficiency, enabling rapid simulations and data fitting on CPU and GPU hardware. While designed for microstructural dMRI, MATI's generalized framework can be extended to other imaging methods, making it a flexible and scalable tool for quantitative MRI research. By enhancing accessibility and efficiency, MATI offers a significant step toward translating advanced imaging techniques into clinical applications.

# 1 Introduction

There is a growing interest in advancing open science and transparency in MRI research (1). These efforts not only help avoid redundant redevelopment of published methods, which can accelerate clinical translation but also enable researchers to replicate and validate others' methods, thereby enhancing reproducibility. In diffusion MRI (dMRI), tools can be categorized into two main types: (i) Data analysis toolboxes that fit biophysical models to dMRI data to extract information on tractography or microstructural parameters non-invasively, including FSL (2), MRTrix3 (3), DSI-Studio (https://dsi-studio.labsolver.org/citation.html), Dipy (4), Dmipy (5), and ACID {David, 2024 #38}; and (ii) Computer simulation toolboxes that use Monte Carlo, Finite Difference, or Finite Element methods at the microstructural level to examine how diffusion pulse sequences and microstructural parameters affect dMRI signals, such as MISST (6), SpinDoctor (7), RMS (8). A few toolboxes, like Camino (9), offer both data fitting and simulation capabilities.

Most of the toolboxes mentioned above focus on neuroimaging and conventional pulsed gradient spin echo (PGSE) acquisitions. Recently, however, there has been a growing interest in applying oscillating gradient spin echo (OGSE) acquisitions and developing quantitative methods for mapping microstructural information at the cellular level, such as MR cell size imaging in cancer (10,11), particularly the clinically feasible IMPULSED method (12). This approach has been quickly implemented in clinical studies across various cancer types, including prostate (13), brain (14), and breast (15) cancer. However, there are currently no open-source toolboxes that comprehensively handle both computer simulations and data fitting for OGSE and MR cell size imaging. Without standardized tools, different groups may implement the same methods differently (e.g., using various fitting techniques), which hampers direct comparison of results across sites and poses challenges for evaluating reproducibility across vendors and healthcare providers. Moreover, even when studies make their core code open-source, implementing this code in specific research can require significant time, effort, and technical expertise. This poses a particular challenge for clinical end users, who may lack the strong technical support needed to adopt and adapt these methods effectively.

This work aims to address this gap by providing an open-source toolbox for microstructural dMRI that offers flexibility in hardware specifications, acquisition parameters, and fitting strategies. The toolbox is designed as a generalized, user-friendly framework with high extensibility, applicable not only to dMRI but also to other quantitative MRI methods. In this framework, computer simulations are broadly defined as generating MRI signals from a microstructure using a specific pulse sequence, while data fitting is framed as fitting a biophysical model—using either analytical equations or simulated dictionaries (16) —to MRI data to estimate microstructural parameters. Despite the diversity of research directions, this generalization holds, enabling the creation of a modular framework where users can "plug in" customized components, such as different microstructures, pulse sequences, biophysical models, or fitting methods. This approach is expected to significantly reduce the time and effort required to develop new MRI methods.

To address these needs, we propose an open-source, MATLAB-based toolbox called MATI (Microstructural Analysis of Tissues by Imaging), designed as a GPU-accelerated microstructural

MRI simulation and data fitting tool with a user-friendly graphical interface (GUI). MATI enables Finite Difference (FD) simulations on arbitrary microstructures with customizable diffusion gradient waveforms, offering extensive flexibility for simulating the effects of diffusion on MRI signals. Additionally, MATI can perform data fitting to extract key microstructural parameters, including cell size, cell density, intra- and extracellular diffusivities, and the transcytolemmal water exchange rate constant. Both simulations and data fitting are GPU-accelerated and accessible via a simple GUI, making the tool suitable for end users with limited research experience. MATI aims to be a valuable resource for the diffusion MRI field, supporting both pre-clinical and clinical research applications.

## 2 Materials and Methods

### 2.1 Overview of MATI

MATI is developed in MATLAB (MathWorks, Natick, MA) and can be conveniently installed as a MATLAB App, with an installation guide provided in the supplemental document. Built on object-oriented programming (OOP), MATI defines key components as classes—such as *ImageData*, *SignalModel*, *PulseSequence*, *FitPars* (fit parameters), *TissueStructure*, and *Simulator*, as shown in Figure 1. This structure offers users high flexibility to customize subclasses for their specific research projects without the need to rewrite simulation and fitting code. Additionally, MATI's built-in class methods automatically validate parameters and handle complex calculations. For instance, a *PulseSequence* object calculates diffusion time $t_{diff}$ from the input diffusion gradient shape, duration $\delta$, and separation $\Delta$, while also ensuring the consistency of all pulse sequence properties. Similarly, a *FitPars* object, when enabled, automatically computes cell density and cellularity if cell size $d$ and intracellular volume fraction $v_{in}$ are fitted, and calculates the diffusion dispersion rate (DDR) (17) from ADCs with multiple diffusion times. These internal functions not only reduce user effort significantly but also minimize the potential for coding errors.

Additional supporting materials are provided with MATI, including: **Physics**, which contains analytical equations for biophysical models; **Tools**, a collection of programming, visualization, and calculation utilities; **External**, which includes third-party software packages for tasks such as reading/writing image files and noise removal (18); and **Examples**, a set of example MATLAB scripts demonstrating specific applications of MATI. While a user-friendly GUI is available for basic operations, advanced users can access the full functionality of MATI through scripting. Additionally, MATIpy, a Python version of the MATI fitting pipeline, is provided in a command-line format for easier integration with other software or workflows.

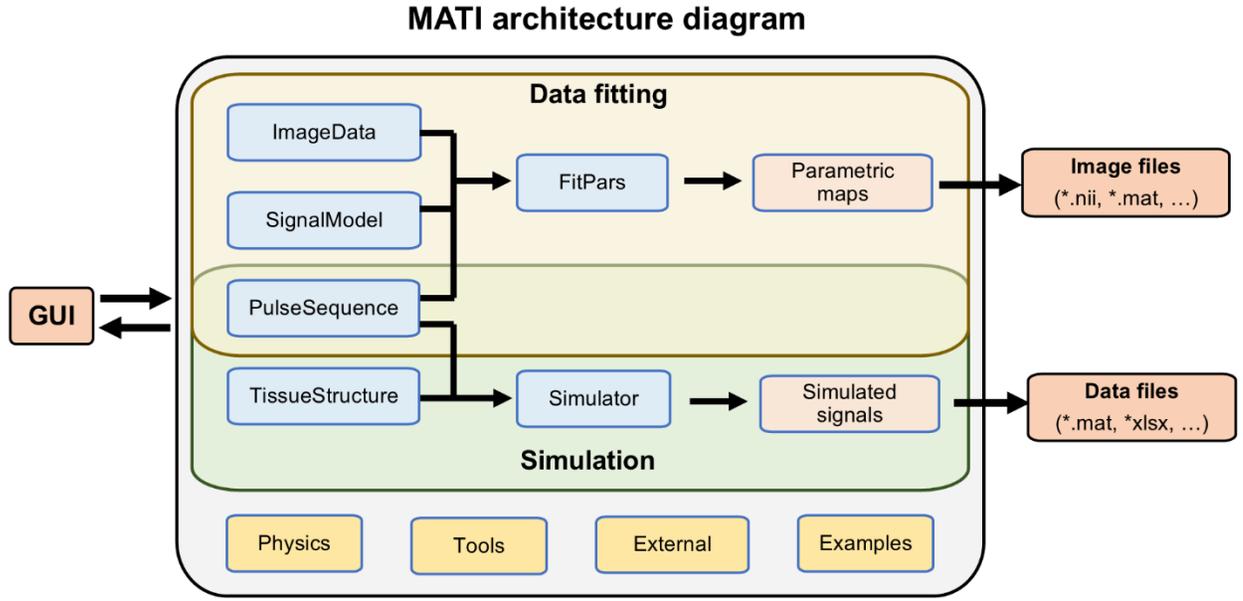

Figure 1 The architecture diagram of the MATI package. The core package includes three main components: simulation, data fitting, and supporting materials. A graphical user interface (GUI) provides easy and friendly operations for standard studies, while examples of MATLAB scripts are provided for advanced usage.

## 2.2 Simulation

### 2.2.1 Simulation pipeline

The MATI simulator can accept arbitrary discretized tissue structures and pulse sequences, as illustrated in Figure 2. Tissue models can either be generated using MATI's tissue generator or imported from segmented histology images. Example scripts are provided for generating 1D, 2D, or 3D arrays of regularly or randomly packed spheres or cylinders, with customizable parameters for cell/axon sizes, density, and cross-membrane permeability. Additionally, third-party software can be used for image auto-segmentation (19), allowing users to convert segmented images into tissue structures compatible with the simulator.

MATI employs an improved FD method to simulate dMRI signals for arbitrary tissue structures (20). The core equation is $\mathbf{M^{n+1}} = (\mathbf{A_0} + A_{\pm x}^n + A_{\pm y}^n + A_{\pm z}^n) \times \mathbf{M^n}$, where $\mathbf{M^n}$ is a column vector representing the magnetization of all tissue grid points at the n$^{th}$ time step. Here $\mathbf{A_0}$ is the FD matrix that describes water diffusion among non-boundary grid points in each time step, while $A_{\pm x}^n$, $A_{\pm y}^n$, and $A_{\pm z}^n$ are FD matrices associated with boundary grid points. Unlike typical FD diffusion simulations, the matrices $A_{\pm x,y,z}^n$ vary with time, as MATI applies a revised periodic boundary condition (RPBC) to reconcile the linearity of diffusion gradients with the periodicity of tissue boundaries. This method modulates the magnetization at boundary points with a phase based on the integral of time-varying diffusion gradients, resulting in time-dependent FD matrices.

Although this approach significantly increases the computational load, it enables a more accurate simulation of dMRI signals. Detailed information about the FD simulation algorithm is provided in reference (20).

Simulated dMRI signals can be visualized and further analyzed, for example, through data fitting using quantitative microstructural methods or ADC spectrum analysis over varying diffusion times, as illustrated in Figure 2.

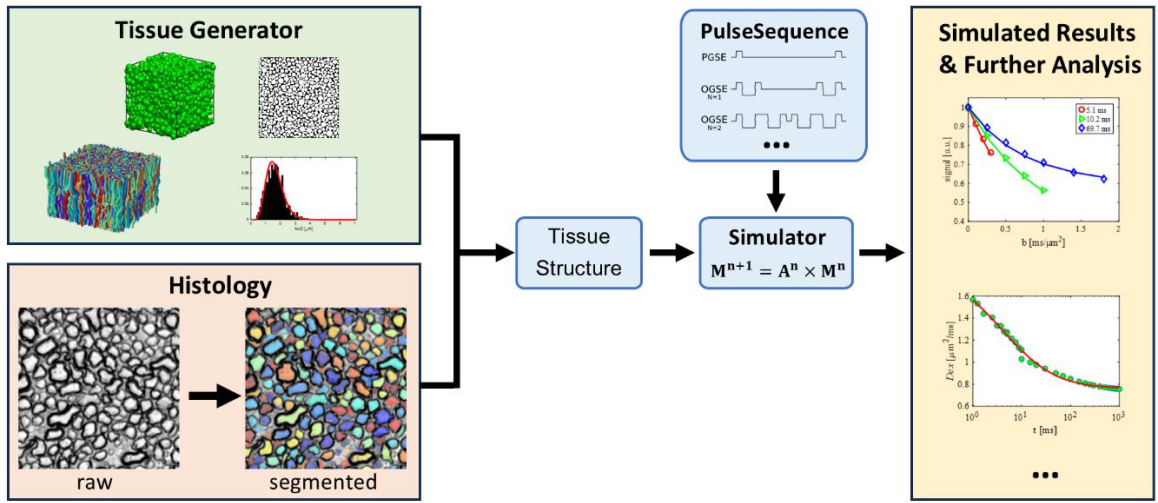

Figure 2 Diagram of the MATI simulation pipeline. The input tissue can be either generated by a tissue generator or segmented from a histology image and then converted to a tissue structure object. The tissue structure can be combined with any pulse sequence to be fed into the simulator. The simulated results can be further analyzed and virtualized using MATI-provided tools.

### 2.2.2 Accelerating simulation

The FD simulation method utilized in MATI was initially developed in C/C++ with MPI (Message Passing Interface) for parallel computing on multiple cores (20). Since the primary operation is simple matrix multiplication, the FD simulation remains highly efficient when implemented in MATLAB. However, to further enhance performance, we propose three acceleration strategies within MATI:

1. **Initialization of FD Matrices**: FD matrices are exceptionally large, and each element must be individually computed due to the heterogeneous distribution of tissue properties (e.g., diffusivity, proton density, permeability, T2). Consequently, calculating these matrices can be time-intensive before simulations begin. By leveraging topological graph theory, the initial FD matrices can be represented as graphs, with diffusion between points treated as directed, weighted edges. MATLAB's built-in functions for handling graphs with directed edges and weighted adjacency matrices can then be utilized, significantly reducing computation time.
2. **Updating FD Matrices**: The primary equation for updating matrices is $\mathbf{M}^{n+1} =$

$\mathbf{A}^n \times \mathbf{M}^n = (\mathbf{A_0} + A^n_{\pm x} + A^n_{\pm y} + A^n_{\pm z}) \times \mathbf{M}^n$. When time-varying diffusion gradients such as OGSE or PGSE with finite durations are applied, FD matrices require updates at each time step where the diffusion gradient is non-zero. We propose two methods for updating FD matrices:

- Method#1: Update the entire large sparse matrix $\mathbf{A}^n = (\mathbf{A_0} + A^n_{\pm x} + A^n_{\pm y} + A^n_{\pm z})$ first, then compute $\mathbf{M}^{n+1} = \mathbf{A}^n \times \mathbf{M}^n$.
- Method#2: Update the smaller sparse matrices $A^n_{\pm x,y,z}$ individually and then calculate $\mathbf{M}^{n+1} = \mathbf{A_0} \times \mathbf{M}^n + A^n_{\pm x} \times \mathbf{M}^n + A^n_{\pm y} \times \mathbf{M}^n + A^n_{\pm z} \times \mathbf{M}^n$.

Although both methods are mathematically equivalent, they differ in programming efficiency. Method #2 updates much smaller sparse matrices, allowing faster updates at each time step, ultimately accelerating FD simulations compared to Method #1.

3. **Utilization of GPU**: Since the FD simulation exclusively uses vectorized matrix operations, it can be efficiently adapted to GPU computation in MATLAB with minimal code modification by employing gpuArray(). This GPU implementation further enhances simulation speed.

## 2.3 Data fitting

### 2.3.1 Data fitting pipeline

MATI establishes a generalized data fitting framework, as illustrated in Figure 3. This framework integrates three primary objects—*ImageData*, *SignalModel*, and *PulseSequence*—into the *FitPars* object, allowing users to select various fitting options, such as denoising, IVIM effect removal, and specific fitting methods.

By default, MATI includes *SignalModel* subclasses for the IMPULSED (12), MRI cytometry (21), and JOINT (22) methods. Users can also define custom *SignalModel* subclasses by specifying fitting parameter labels and a signal function, which can then be seamlessly integrated into MATI. This plug-and-play capability enables users to leverage the entire data fitting framework without rewriting fitting code. The framework's flexibility allows it to accommodate any biophysical model with a signal function, making it extendable to other quantitative MRI methods. To illustrate this adaptability, MATI provides an example of a non-diffusion *MT* (quantitative magnetization transfer) as a *SignalModel* subclass for fitting SIR (selective inversion recovery) qMT (23).

Each *SignalModel* object encapsulates a signal function and default fitting options, giving model creators control over the recommended fitting strategies. Users, however, can override these defaults in the *FitPars* object to suit their needs. For instance, the default settings for the "IMPULSED_vin_d_Dex" model include dictionary matching fitting, denoising, IVIM effect removal, time-dependent ADC calculations, and GPU usage if available. Users can flexibly adjust any of these options. Additionally, derived parameters, such as 3D cell density and 2D cellularity (23), are computed from the fitted $d$ and $v_{in}$ values. The ADC change, $\Delta\text{ADC} = [\text{ADC(OGSE)} - \text{ADC(PGSE)}]$ (24) is also calculated when both ADC(OGSE) and ADC(PGSE) are available.

MATI features an interactive GUI for pixel-wise assessment of fitting quality. Users can select any pixel within the VOI (volume of interest) in the left figure to display the raw dMRI signals in the middle figure. The right figure then shows the processed dMRI signals (e.g., normalized after IVIM effect removal), fitted parameters, and model-predicted dMRI signals. This interface enables users to efficiently evaluate the fitting quality for individual pixels.

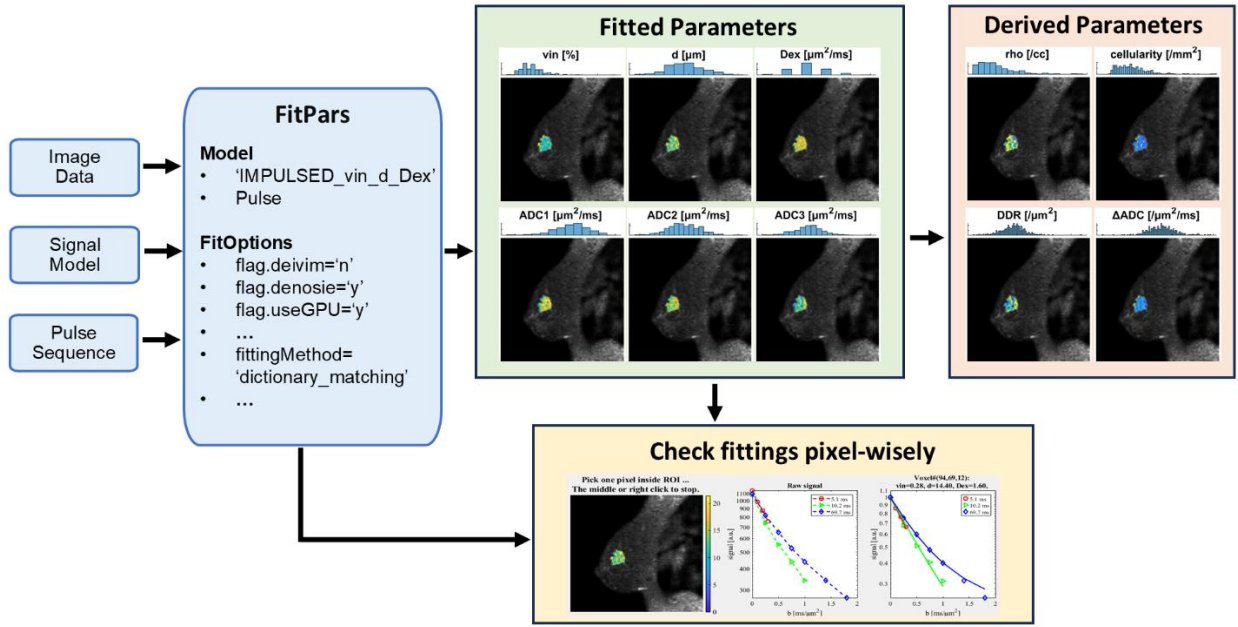

Figure 3 Diagram of the MATI fitting pipeline. The MATI framework integrates three primary objects—*ImageData*, *SignalModel*, and *PulseSequence*—into the *FitPars* object, allowing users to select various fitting options, such as denoising, IVIM effect removal, and specific fitting methods. In addition to fitted parameters directly from fittings, other derived parameters will also be calculated such as deriving effective cellularity from $v_{in}$ and $d$. Optionally, an interactive GUI allows users to check data fittings at any pixel such as raw signals, normalized signals after removing the IVIM effect, and predicted signals by the fitted model. The fitting was performed on a breast cancer patient.

### 2.3.2 Fitting methods

As a generalized fitting framework, MATI allows for the flexible incorporation of various fitting methods. The following methods are available by default:

1. **NLLS (Non-Linear Least Squares)**: As a traditional method widely used for decades, NLLS minimizes a cost function to determine fitted parameters on a per-pixel basis (12). MATI supports both local and global optimization algorithms, with options for different MATLAB solvers and multiple starting points.
2. **GR (Bayesian Approach with Grid Search)**: Based on Bayes' theorem, this approach leverages prior parameter probability to achieve accurate and robust data fitting. MATI combines Bayesian grid search with maximum likelihood to estimate parameter probabilities, as described in (25).

3. **PR (Supervised Machine Learning Polynomial Regression)**: A data-driven method, polynomial regression learns the mapping from noisy measurements to model parameters without requiring an analytical forward model. This approach has been previously used for dMRI data fitting (26).
4. **DM (Dictionary Matching)**: Commonly applied in MR Fingerprinting (27), this method matches acquired signals to a precomputed dictionary of predicted signals, reducing the risk of local minima and improving accuracy and robustness. MATI also introduces a knowledge-informed DM variant that incorporates parameter probability densities, further enhancing fitting precision.

To demonstrate MATI's flexibility in selecting fitting methods and options, we evaluated the following eight combinations of methods and processing resources:
1. Knowledge-informed DM method using GPU
2. Conventional DM method (without parameter probabilities) using GPU
3. PR method using GPU
4. Knowledge-informed DM method using CPU
5. Conventional DM method using CPU
6. PR method using CPU
7. NLLS method using CPU
8. GR method using CPU

### 2.4 GUI

MATI features a user-friendly GUI that simplifies running simulations and data fitting. The interface supports tasks such as loading data and masks, drawing and verifying masks, selecting models and pulse sequences, configuring fitting options, visualizing results, and exporting outcomes to MATLAB .mat files or NIfTI image files.

For convenience, MATI includes several published pulse sequences, such as:
- **IMPULSED** sequences for human breast cancer with a single-axis gradient strength of 65 mT/m (12), for human prostate cancer with 45 mT/m (13), and for animal tumor studies with 360 mT/m (10).
- **VERDICT** for prostate cancer (28).
- **SSIFT** for brain cancer (29).

The supplemental document includes examples with screenshots to guide users in performing simulations and data fitting using the MATI GUI.

### 3 Results

All simulations and data fittings were conducted on a 24-core CPU processor (Intel i9-14900, 2000 MHz) with 64 GB of memory and a GPU (NVIDIA GeForce RTX 4070).

Figure 4 shows the acceleration achieved by three new approaches in MATI simulation with varying tissue size. An OGSE sequence was simulated with a trapezoidal cosine-modulated gradient waveform using parameters: number of cycles N = 1, b = 1000 s/mm$^2$, duration $\delta$ = 40 ms, separation $\Delta$ = 51.4 ms, and TE = 110 ms. The simulated tissue model consisted of a tightly-packed spheres with a membrane permeability of 0.024 µm/ms. Figure 4A demonstrates that the topologic graph theory method significantly reduces initialization time for FD matrices. —from hours to seconds—yielding a 4-order-of-magnitude acceleration. Figure 4B shows that different FD matrix updating methods substantially impact computation time. Method #2, which avoids updating the large sparse matrix, achieves a 4-fold speedup. Figure 4C illustrates a 3-fold acceleration when using the GPU compared to the CPU alone.

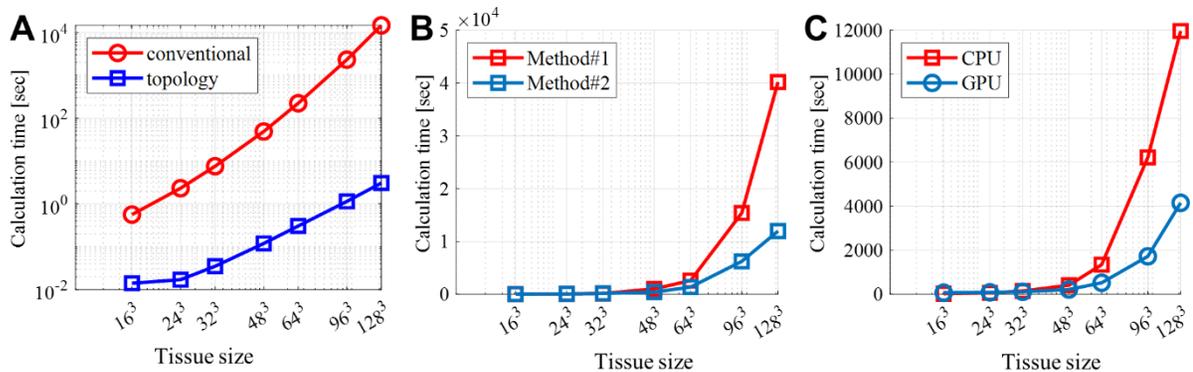

Figure 4 Acceleration of new approaches for the MATI simulation. (A) Calculation time for generating the initial FD matrix. (B) Comparison of two methods to update the FD matrix with time-varying OGSE diffusion gradients. (C) GPU acceleration with respective to tissue size.

Figure 5 presents a comparison of eight different combinations of fitting methods and options, using simulated data across 100,000 pixels under the IMPULSED protocol with random Rician noise at SNR=20, ground truth $d$ =15 µm and $v_{in}$=0.7. Figure 5A highlights that the DM and PR methods are significantly faster than NLLS and GR, achieving two orders of magnitude acceleration on the CPU. With GPU support, DM and PR methods achieve three orders of magnitude acceleration, fitting 100,000 pixels in under 5 seconds. Figure 5B displays the mean and standard deviation (STD) of the fitted cell size $d$ for each of the eight combinations, indicating accuracy and precision. Despite its high speed, the PR method shows the largest deviation from the ground truth. While NLLS and GR provide good accuracy, they have relatively low precision. The knowledge-informed DM method outperforms all others, offering the highest accuracy and precision alongside minimal computation time. Consequently, the knowledge-informed DM method is recommended as the preferred fitting method in MATI.

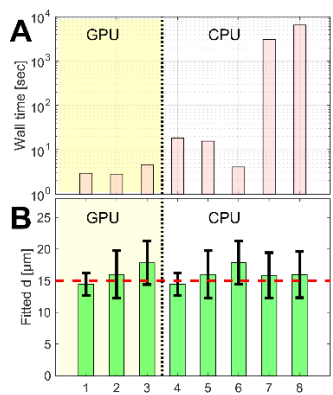

Figure 5 Comparison of 8 different combinations of fitting methods and options, as explained in the main text. The data were simulated in 100,000 pixels using the IMPULSED acquisition protocol with random Rician noise at SNR=20 and a ground truth $d$ = 15 μm and $v_{in}$ = 0.7. The total fitting time was compared in (A) and the mean and STD of fitted $d$ are summarized in (B). The shaded area indicated GPU was used in 1, 2, and 3, while the rest used CPU only.

## 4   Discussion

MATI is a unique microstructural dMRI toolbox offering both simulation and data fitting functionalities, making it a comprehensive resource for microstructural dMRI research. As a standalone MATLAB app with a GUI, MATI is easy to install and use, requiring no programming experience. This accessibility has the potential to facilitate the adoption of advanced imaging methods in clinical studies.

Beyond user-friendliness, MATLAB's optimization capabilities enhance MATI's computational efficiency. Although some studies suggest that languages like Julia may outperform MATLAB for specific tasks like non-linear least squares fitting (30), MATI leverages highly vectorized matrix operations. By using high-performance numerical libraries such as LAPACK for linear algebra and matrix computations, MATI achieves remarkable efficiency, with fitting times reduced to just a few seconds for 100,000 pixels. This level of performance further supports MATI's applicability in clinical research.

While MATI is primarily designed for microstructural dMRI, it incorporates a generalized framework that can extend to other imaging methods. For example, the FD simulation algorithm in MATI has been applied to simulate diffusion effects in dynamic susceptibility contrast (DSC) (31) and T1ρ (32) MRI. Any quantitative biophysical model paired with experimental data can utilize MATI's flexible framework, eliminating the need for redundant coding. As a demonstration of this versatility, MATI includes data fitting for selective inversion recovery (SIR) in quantitative magnetization transfer (qMT) (33).

## 5   Conclusion

MATI is a user-friendly, efficient toolbox that combines simulation and data fitting for microstructural dMRI. Its GUI-based MATLAB app allows researchers without programming expertise to perform advanced analyses, supporting the clinical translation of imaging techniques.

Optimized for high-performance computing, MATI handles large-scale data fitting tasks efficiently. While designed for microstructural dMRI, its flexible framework can be extended to other imaging methods, demonstrating MATI's potential as a versatile tool for quantitative MRI research and clinical applications.

## Acknowledgments

This work was funded by NIH grants UH3CA232820, R01CA269620, R21CA270731, P30CA068485, and R01DK135950.

# MATI: A GPU-Accelerated Toolbox for Microstructural Diffusion MRI Simulation and Data Fitting with a User-Friendly GUI

A Quick User Guide

## Contents

# 1 Installation

MATI is distributed as a MATLAB App. The installation file *MATI.mlappinstall* can be downloaded from https://github.com/jzxu0622/mati.

The installation of simple and straightforward as shown in Figure 1. Choose MATLAB -→ the **APPS** tab -→ Click **Install App** -→ choose the downloaded MATI.mlappinstall to install.

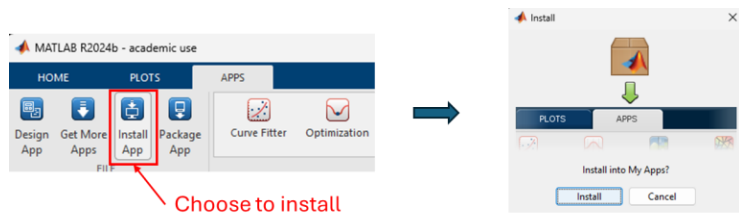

Figure 1 Installation of MATI

After the installation, MATI will appear in the MATLAB APPS list. If you hover the mouse over the MATI button, you can find the installation information as shown in Figure 2. Note that the **File Location** is where the MATI App is installed. If advanced users want to use MATI script programming, they need to either (1) add the File Location to the MATLAB search path or (2) keep the MATI App open when running scripts. The purpose is to ensure the namespace **+mati** in the MATI package is in MATLAB's search path.

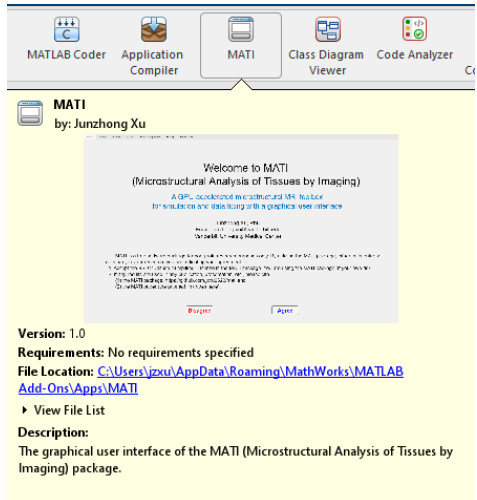

Figure 2 Installation information on the installed MATI. The File Location is where MATI was installed. It needs to be added to the MATLAB search path when using scripts that call the MATI package.

NOTE: Some incompatibility was found between different MATLAB versions. For example, MATI installed on MATLAB 2023a has issues displaying switch buttons correctly in MATI GUI. It is highly recommended to install MATI to the corresponding MATLAB version. The current version is MATLAB 2024b.

## 2   Usage

MATI GUI can be opened by simply clicking the MATI button in the MATLAB APPS list. MATI organizes different functionalities into different Tabs as shown in Figure 3. You cannot switch freely between Tabs because some Tabs require additional information from previous Tabs. MATI GUI will guide you through the necessary Tabs. You must click **Agree** to continue.

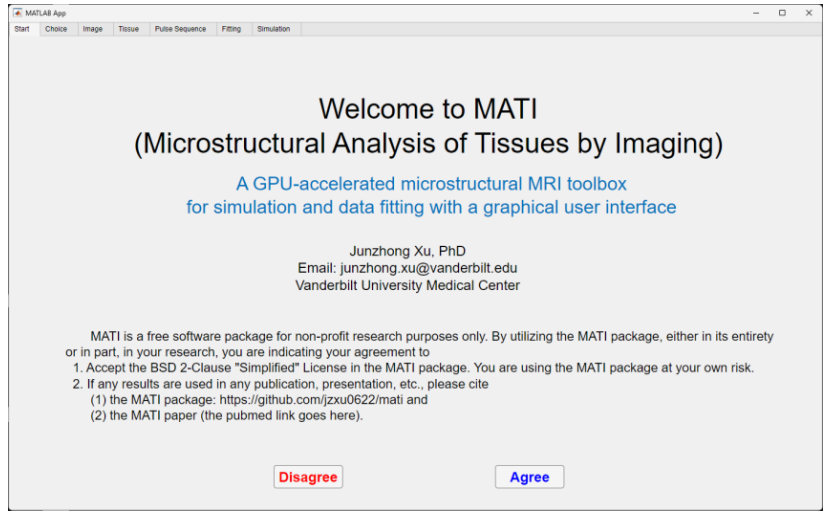

Figure 3 The start page of MATI.

On the next page, you have two options to use MATI: Simulate dMRI signals or Fit dMRI Images. First of all, it is highly recommended to choose your working directory. By default, your working directory is your MATI installation location, as shown in Figure 4. It is recommended to change your working directory to the folder that contains your data, which makes it convenient to choose both input data files and the output folder.

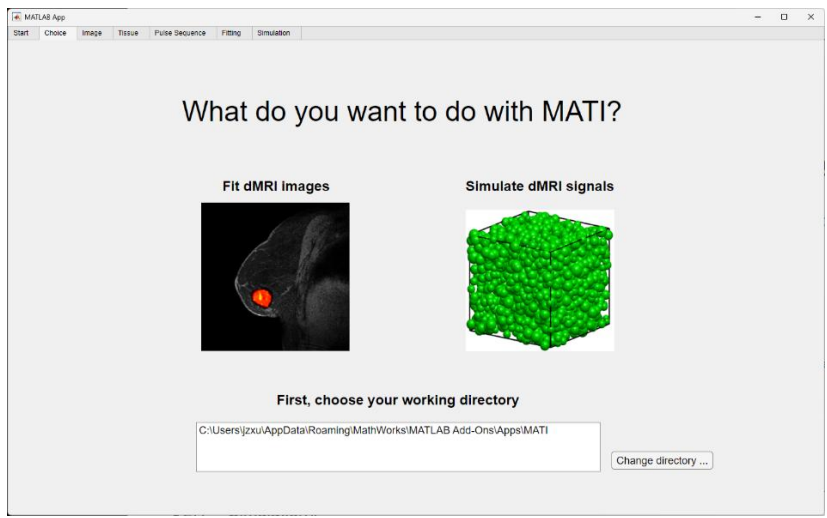

Figure 4 Options to use MATI and to choose the working directory.

Next, we will use two examples to demonstrate how to use MATI for computer simulations and data fitting.

## 2.1 Simulation using GUI

After clicking the packed-sphere tissue model figure on the right in Figure 4, you will be brought to the page to load tissue structure for computer simulations. In the dropdown list, you can find some pre-generated tissue structures in Figure 5. These files contain usually simple structures such as free water, regularly packed cylinders or spheres. Users can choose "**User_defined**" to load their own pre-generated tissue structure. There are two ways to generate user-customized tissue structures:

1. MATI provides example scripts to generate 1D, 2D, or 3D tissue structures.
2. Users can take histology images, perform segmentation to separate intra- and extra-cellular spaces and myelin (if any), and then convert segmented images to tissue structures.

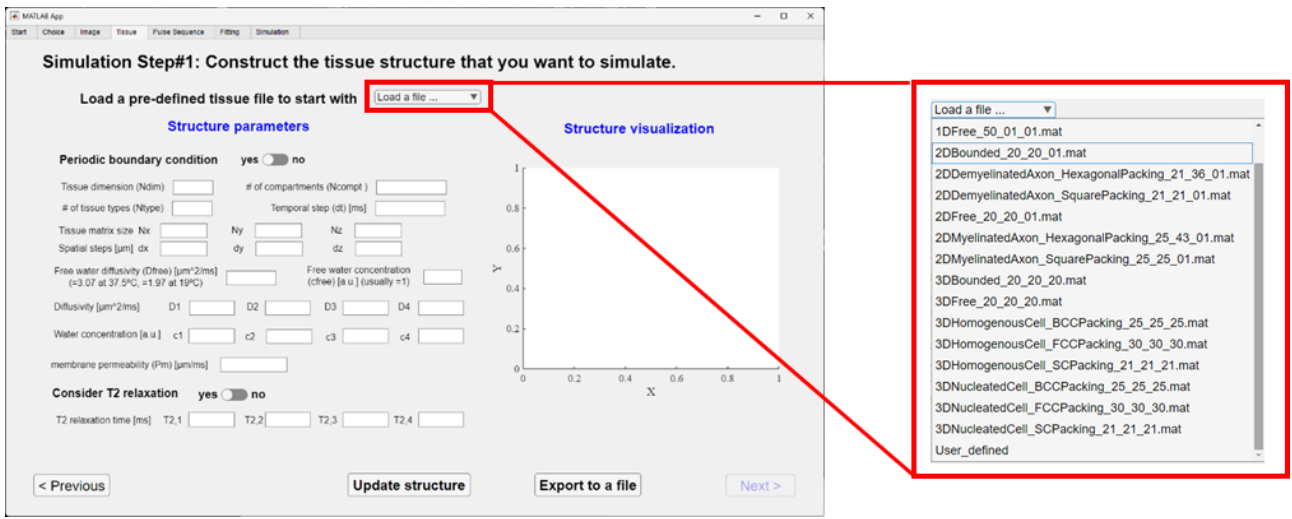

Figure 5 Tissue page to load pre-generated modeled tissue structure. Some tissue structures are provided by default on the tissue file list but the users can load their pre-generate tissue structures.

After loading the tissue structure, all tissue-related parameters will show up and the tissue will be visualized on the right. Different colors/values indicate different tissue types. Their corresponding values such as spatial steps ($dx$, $dy$, $dz$), water concentration and diffusion coefficients of each component, and cross-membrane permeability can be modified by users. After clicking the "**Update structure**" button, all the page including the structure visualization will be refreshed. If users want to keep their modifications for future use, one can export the current tissue structure to a MATLAB *.mat file to be loaded in the future.

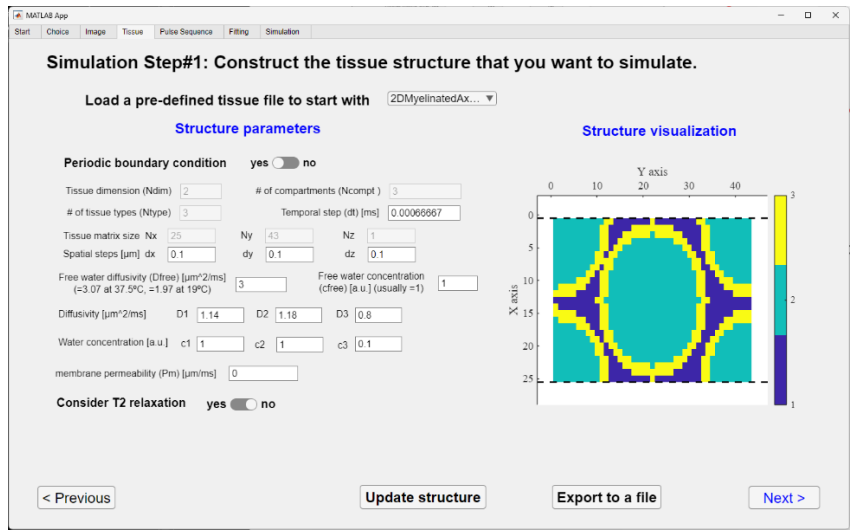

Figure 6 Display of tissue structure parameters and visualization of the tissue structure.

NOTE:
1. Tissue structure can be either 1D, 2D, or 3D.
2. A tissue type is defined as a type of tissue with a unique tissue property such as water concentration, diffusivity, T2, etc. For example, the tissue shown in Figure 6 has three tissue types: extra-axonal (index =1), intra-axonal (index=2), and myelin (index=3).
3. A tissue compartment is defined as a bounded region that contains a tissue type. For example, many axons are the same tissue type but different tissue compartments. Such a definition provides flexibility to simulate signals from individual cells/axons.

After clicking the "**Next**" button, you will be brought to the Pulse Sequence Tab as in Figure 7. In the dropdown menu, several previously published pulse sequences are listed. Users can pick any or choose "**User_Defined**" to load their own pre-define pulse sequence.

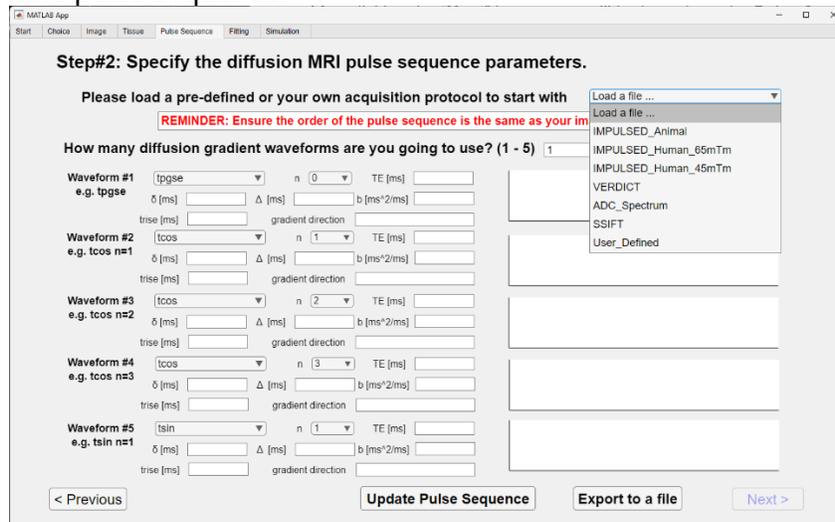
Figure 7 Tab to load pulse sequence.

After a pulse sequence is chosen, all pulse sequence parameters and its gradient waveform will be visualized. For example, Figure 8 shows the pulse sequence parameters and gradient waveforms used in the "IMPULSED_Human_65mTm" protocol. Users can modify any pulse parameters and click "**Update Pulse Sequence**" to refresh the visualization. If users want to use the modified pulse sequences in the future, click "**Export to a file**" so that the pulse sequence will be saved in a MATLAB *.mat file for future use.

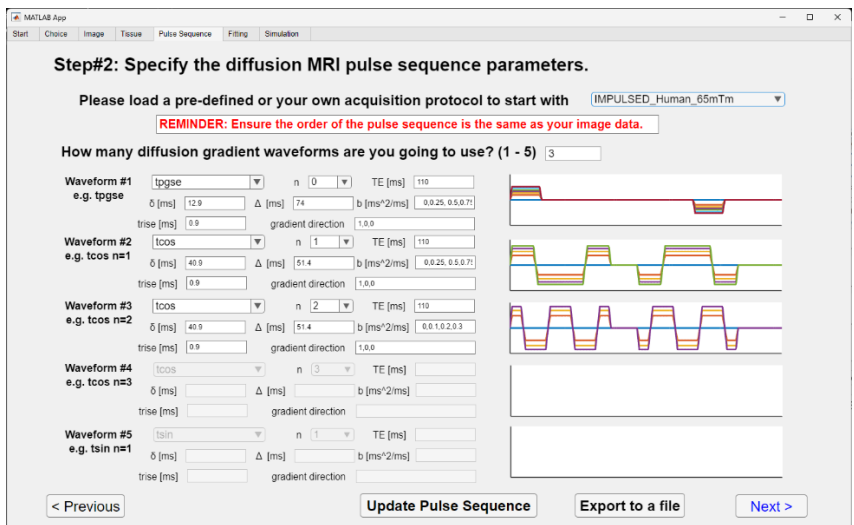
Figure 8 Visualization of the pulse sequence used in the "IMPULSED_Human_65mTm" protocol.

The next is the simulation tab as shown in Figure 9. Users can specify the name of this simulation result file and the folder that the result file will be saved. At this moment, the "**Run simulation**" button is gray because users need to check the parameters first. This is because MATI uses the Finite Difference simulation with the explicit discretization scheme, with which the Courant–Friedrichs–Lewy (CFL) condition must be met. In other words, the spatial and temporal steps cannot be too large otherwise the simulation cannot converge. Moreover, the spatial step should be small enough to capture the spatial oscillation of diffusion gradient-induced phase modulation. All simulation parameters will be checked automatically after clicking "**Check parameters**".

NOTE:
1. MATLAB on Mac OS does not support GPU usage anymore. Therefore, the GPU option can be turned on with Windows and Linux systems only.

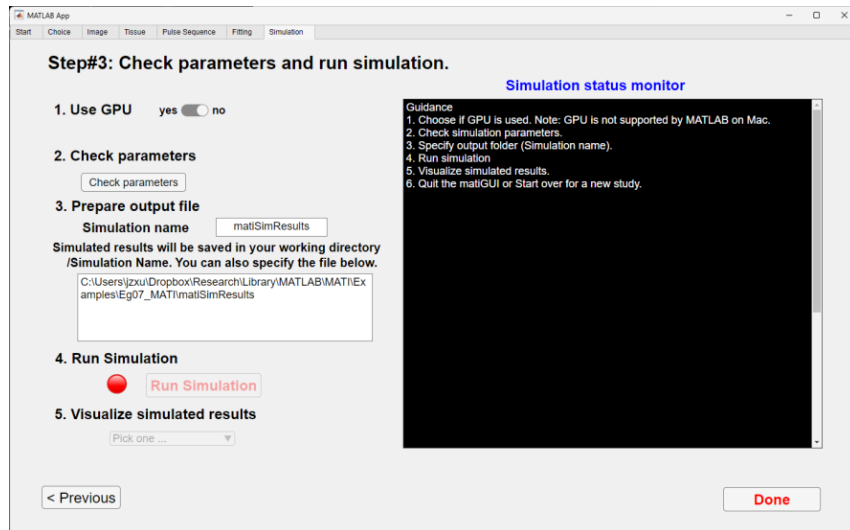

Figure 9 Simulation Tab.

After all parameters are checked successfully, the "**Run simulation**" button becomes blue and the green light next to the button is on (see Figure 10), suggesting the simulation is ready to start.

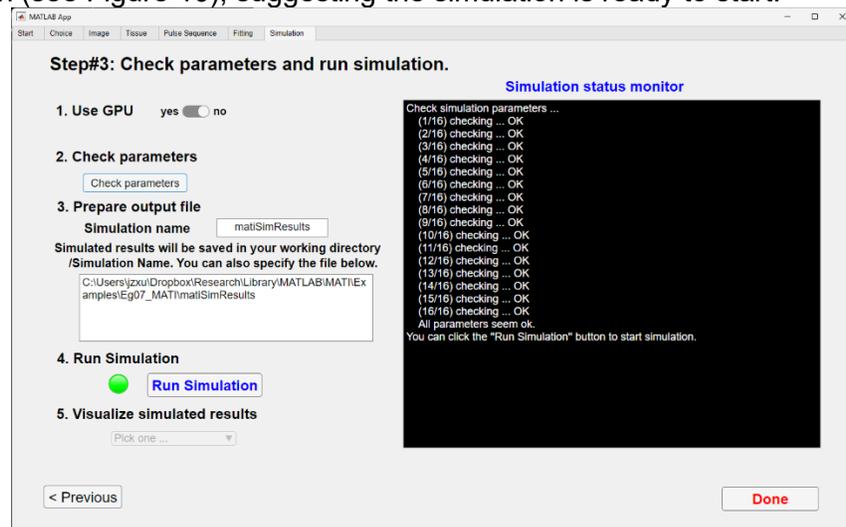

Figure 10 It is ready to start the simulation after all parameters are checked successfully.

After the simulation is done, the simulated results will be saved as a MATLAB *.mat file in the name and folder specified by the user. Moreover, users can choose from the dropdown menu to visualize simulated results quickly as in Figure 11. After everything is done, click "Done" to exit MATI.

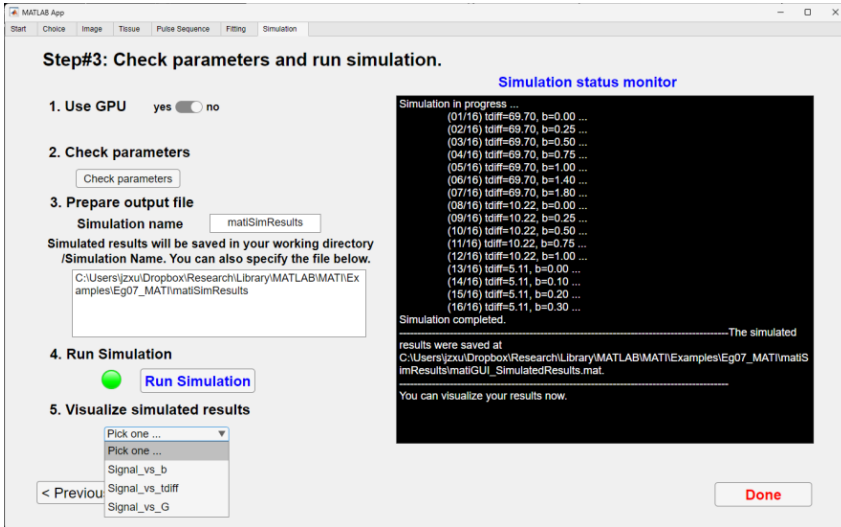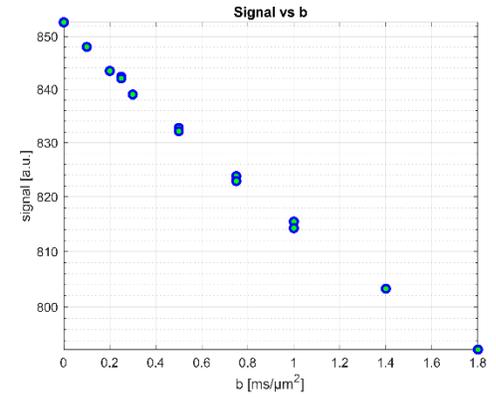

Figure 11 After the simulation is done, users can choose to visualize the simulated results, such as simulated signals vs b.

## 2.2 Data Fitting using GUI

This section demonstrates how to use MATI GUI to run data fitting in a breast cancer patient. After choosing "**Fit dMRI Images**" in Figure 4, users will be brought to the Image Tab (Figure 12). Users have two options:

1. Load a stacked image file, in which all images acquired in different scans (e.g., from different PGSE and OGSE acquisitions) were already stacked into a single image. This is recommended because users can first pre-process these images with co-registration using other third-party software.
2. Users can also load individual raw images from each acquisition. Note that patient movement between different acquisitions may occur and this could impact the fitting results.

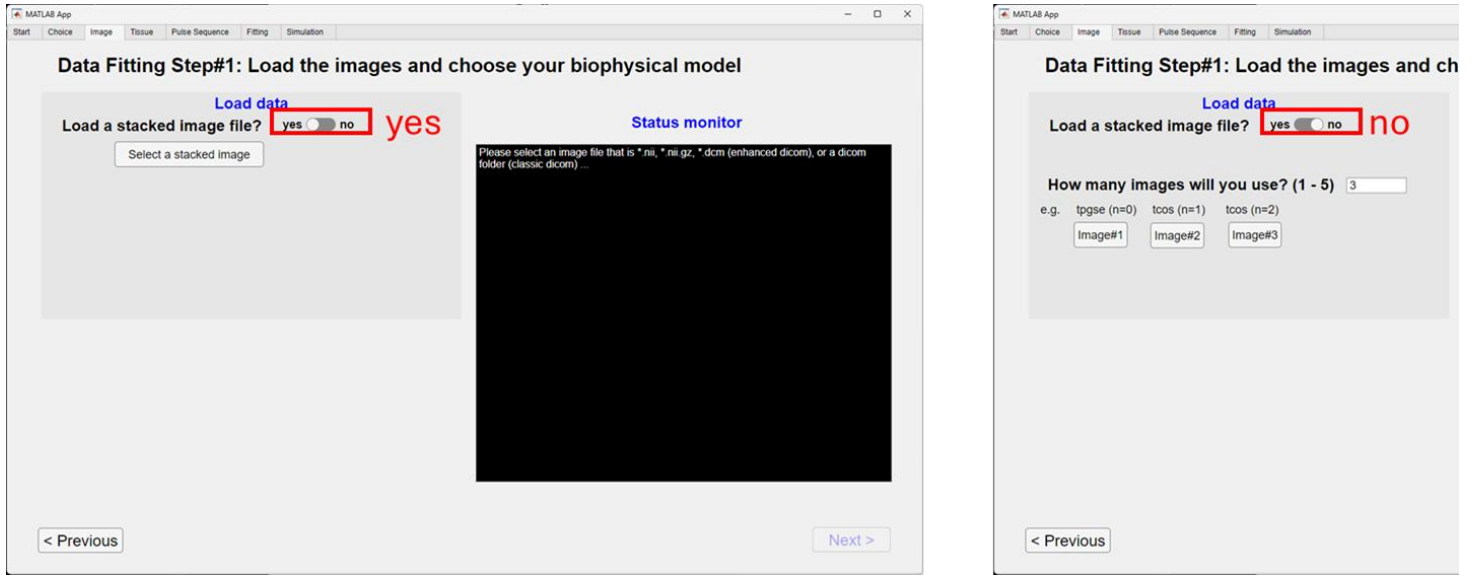

Figure 12 Tab for loading images.

After the image(s) are loaded, users need to load a binary mask image that depicts the volume of interest (VOI) of the lesion(s). There are two options (Figure 13):

1. Load a pre-generated binary mask file; or
2. Manually draw multi-slice binary mask inside MATI.

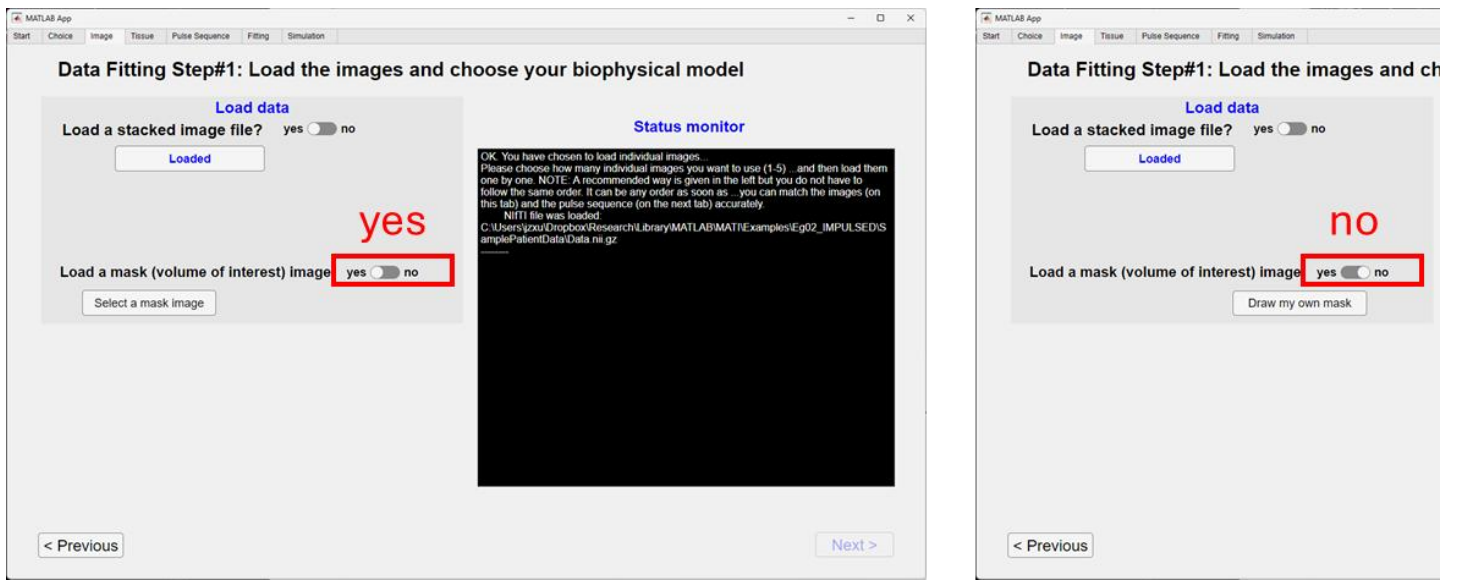

Figure 13 Two options to load or manually draw a binary multi-slice mask.

Here, we assume the mask file is loaded in. Please refer to Section **Error! Reference source not found.** for details about manually draw multi-slice binary mask in MATI. After the mask is loaded or drawn, the following windows will pop up to show the the multi-slice, b=0, T2-weighted image overlaid with the mask (Figure 14). Users can choose Yes to accept the mask or No to draw/redraw the mask.

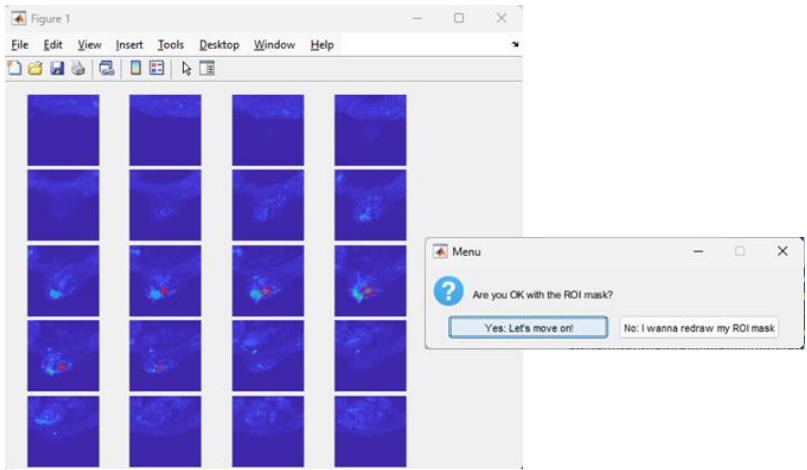

Figure 14 Visualization of the multi-slice, b=0, T2-weighted image overlaid with the mask. Users can choose "Yes" to accept the mask or "No" to draw / redraw the mask.

After the data and mask are ready, users can choose the biophysical model to fit. In addition to the MRI method such as Microstructural dMRI, users can also choose qMT and CEST. For dMRI, the available Signal models include models from the IMPULSED, JOINT, VERDICT, and EXCHANGE models (Figure 15).

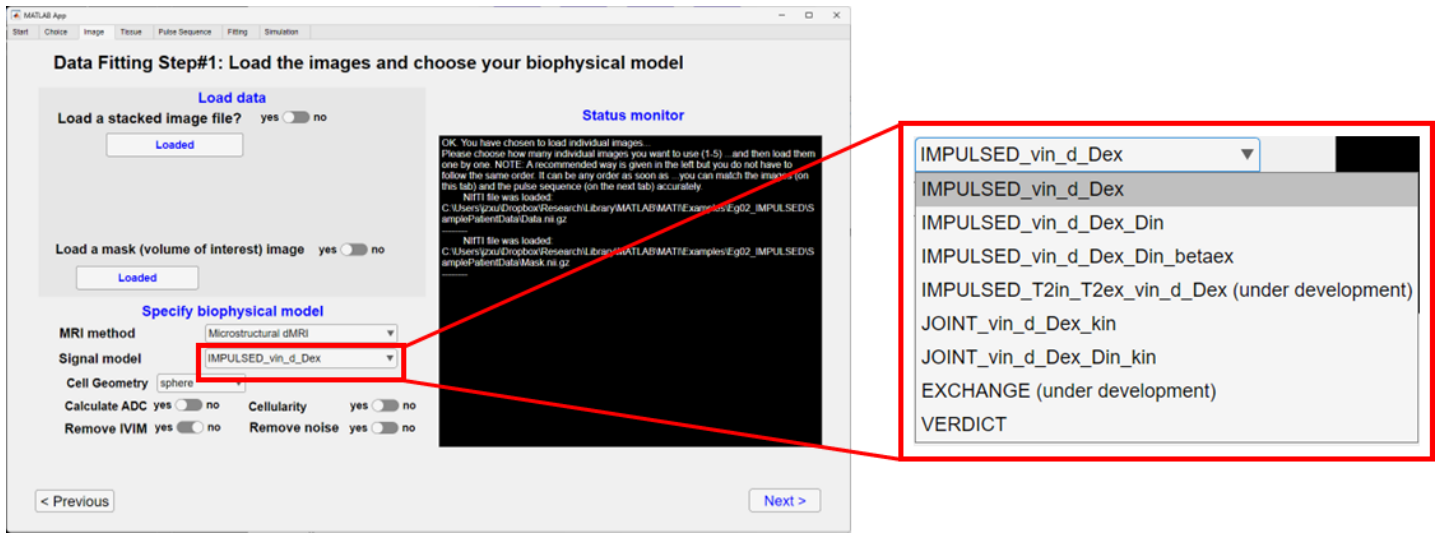

Figure 15 Choices of MRI methods (e.g., dMRI, qMT, CEST) and signal models. The dropdown menu shows the available signal models for the Microstructural dMRI method.

NOTE:
1. Users can choose **Cell Geometry** to "sphere" for fitting cells, while to "cylinders" for fitting straight axons.
2. If "**Calculate ADC**" is "yes", MATI will automatically fit ADC maps at individual diffusion times. The output results are ADC1, ADC2, ..., ADCn, where n indicates the $n^{th}$ diffusion time.
3. If "**Cellularity**" is "yes", the 3D cell density (# of cells in a unit volume) and 2D cellularity (# of cells in a unit area) will be calculated if cell size $d$ and intracellular volume fraction $v_{in}$ are fitted.
4. If "**Remove IVIM**" is "yes", the IVIM effect will be removed. Specifically, a mono-exponential fitting is performed with b values within the Gaussian Phase Approximation, such as in the range of 200 and 1,000 s/mm$^2$. This process is done for signals with the same diffusion time since diffusion time and shape have different influences on dMRI measurements. If "**Remove IVIM**" is "no", a simple normalization will be performed with the b=0 image. This is an important procedure for any dMRI data fitting since different scans (e.g., PGSE, OGSE with n=1, and OGSE with n=2) may have different receiver gains, artificially causing large signal differences between scans with different diffusion times. These normalized signals will be used for data fitting and also included in outcome results.
5. If "**Remove noise**" is "yes", a third-party algorithm by Veraart et al. will be used to remove noise.

Clicking "Next", the users will be brought to the Pulse Sequence Tab. Note that both simulation and data fitting share the same pulse sequence, users can refer to descriptions related to Figure 7 that are mentioned above.

NOTE:
1. It is extremely important to align the image data with the corresponding pulse sequence used in the acquisitions. All images should be a 4D volume ($N_x \times N_y \times N_z \times N_t$) where $N_t$ is the total number of measurements and is equal to pulse.Nacq.

Clicking "Next", the users will be brought to the Pulse Sequence Tab. There are several established fitting methods as shown in Figure 16. The Dictionary Matching fitting method is recommended since it is fast (e.g., a few seconds for most tumor VOIs), accurate, and more robust in noisy clinical data.

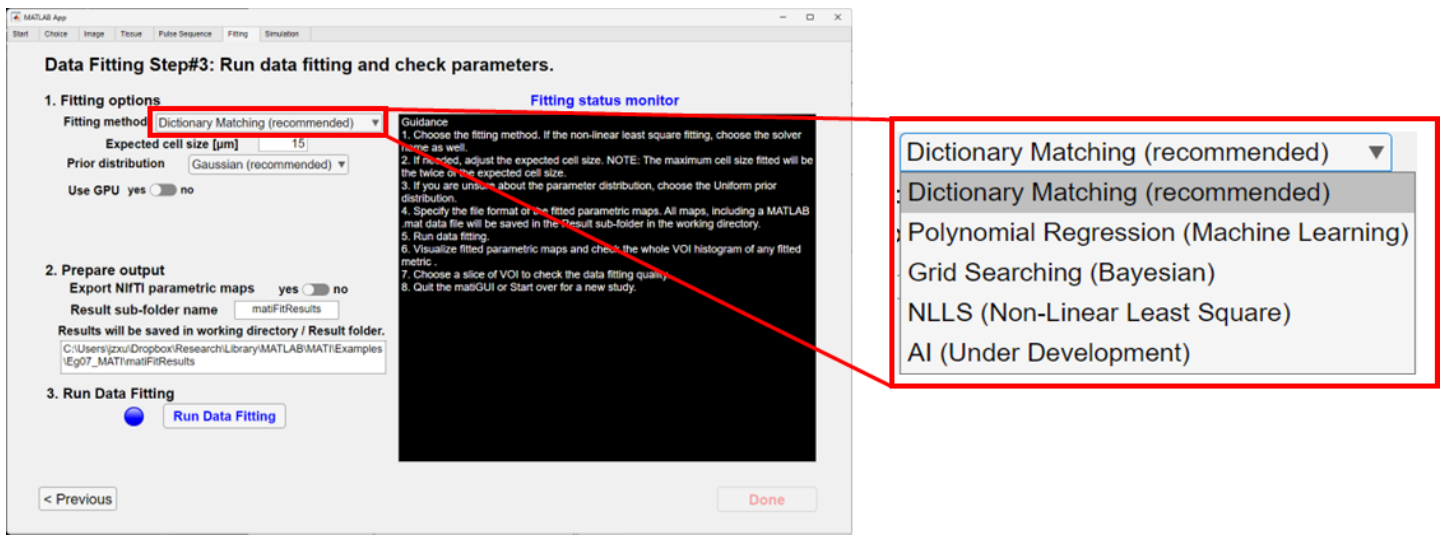

Figure 16 Data fitting Tab.

NOTE:
1. The "**Expected cell size**" is the estimated mean cell size of the tissue that is investigated. If there is no knowledge about this, please still try to set it to a guessed value. For example, typical cancer cells should be 10 – 20 µm while liver cells are usually > 20 µm.
2. Another key feature about "**Expected cell size**" is that the cell size fitting range is twice the expected cell size. For example, if the expected cell size = 15 µm, the cell size fitting range is 0 – 30 µm.
3. The "Prior distribution" indicates the probability distribution of the fitted parameter $d$. It is needed in the Bayesian approach but it can also be combined with the dictionary matching method.
4. Users can choose if all fitted and derived parameters are exported to NIfTI image files. If so, a third-party software NIfTI will be used to save *.nii.gz files.

If users tested MATI with the example data of a breast cancer patient, it should take less than a few seconds for data fitting but it takes some time to export all parametric maps to NIfTI images. After that, the option "**Visualize fitted parametric maps**" appears, with which a dropdown menu shows all fitted and derived parametric maps (Figure 17). The users can pick any of these parameters to visualize the parametric maps as well as the histogram of the parameter in the VOI.

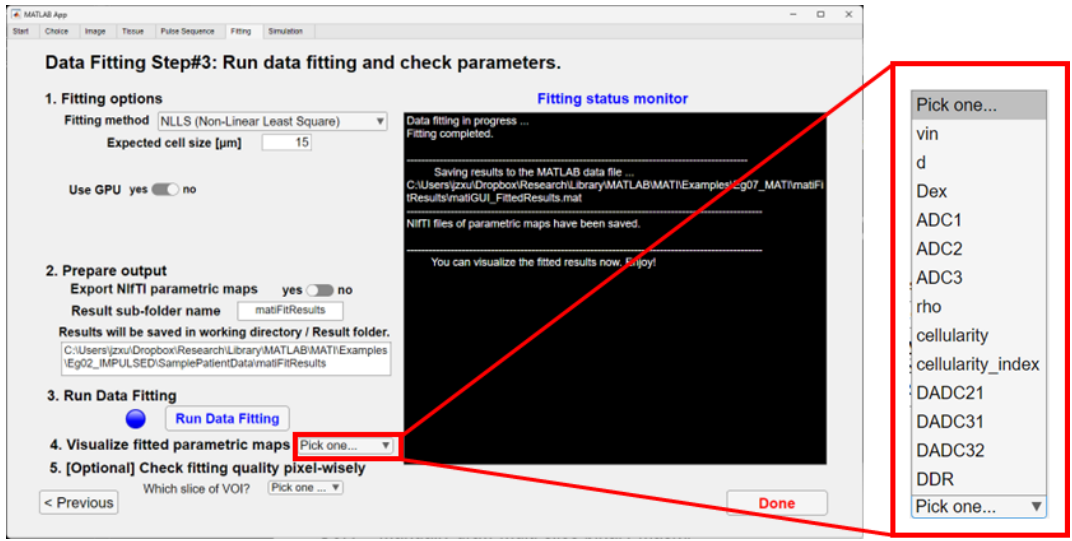

Figure 17 After data fitting is completed, users can choose any parameter to visualize the map and the whole-VOI histogram.

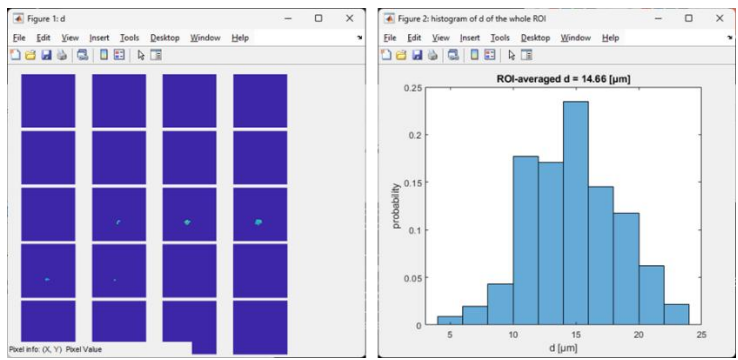

Figure 18 For the fitted cell size $d$, the left shows the multi-slice parametric map and the right shows whole-VOI histogram and VOI-averaged cell size.

Finally, users can choose "**[Optional] Check fitting quality pixel-wisely**". A GUI will pop up so that users can manually pick any pixel to check data fitting quality as shown in Figure 19.

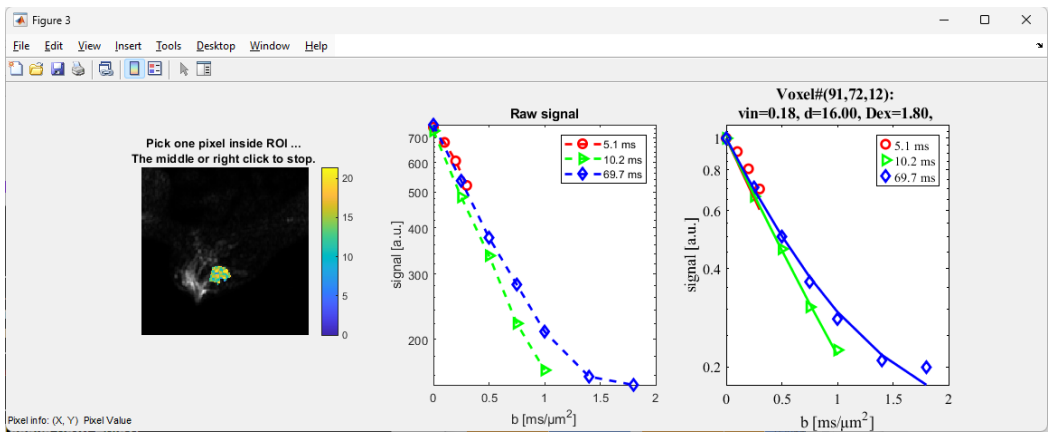

Figure 19 MATI GUI to check fitting quality pixel-wisely. (left) The fitted cell size map overlaid on b=0 T2-weighted image. Users can manually pick any pixel. (middle) The raw image signal at the picked pixel. (right) The processed signals (e.g., normalized) are shown as markers and the fitted model-predicted signals are shown as solid times. Different diffusion times are shown with different colors. All fitted parameters of this pixel are shown in the figure title.